\documentclass[a4paper,12pt]{article}
\usepackage[margin=2.5cm]{geometry}
\usepackage{graphicx, caption, subcaption,lipsum}
\usepackage{amsmath}
\usepackage{amsfonts}
\usepackage{amssymb}
\usepackage{newtxtext} 
\usepackage[slantedGreek]{newtxmath}
\DeclareCaptionFont{myfontsize}{\fontsize{10}{11}\selectfont}
\newcommand{\dd}{\mathrm{d}}
\newcommand{\dr}{\partial}
\newcommand{\ee}{\mathrm{e}}
\newcommand{\qe}{q_{\mathrm{e}}}
\newcommand{\kb}{k_{\mathrm{B}}}

\usepackage{color}

\title{Link between Continuous and Discrete Descriptions of Noise in Nonlinear Resistive Electrical Components}

\author{
	Lucas Désoppi$^1$, Bertrand Reulet$^1$ \\
	\small $^1$Département de physique, Institut Quantique\\
	\small Université de Sherbrooke, Sherbrooke QC, Canada \\
	\small \texttt{lucas.desoppi@usherbrooke.ca} \\
	\small \texttt{bertrand.reulet@usherbrooke.ca}
}

\date{}

\begin{document}
	ICNF-2025 \phantom{xxxxxxxxxxxxxxxxxxxxxxxxxxxxxxxxxxxxxx} June 17-20, 2025 – Taormina, Italy
	
	{\let\newpage\relax\maketitle}
	
	\fontsize{11}{12}\selectfont

	
	
	

	\textbf{Abstract:}
	We consider the modeling of noise in a nonlinear, classical, resistive electrical component using two models: i) a continuous description based on a stochastic differential equation with a white thermal Gaussian noise; ii) a discrete, shot noise model based on a Markovian master equation. We show that thermodynamics imposes in i) the use of the Hänggi-Klimontovich (H-K) prescription when the noise depends on bias voltage, and implies a generalized Johnson-Nyquist relation for the noise where the conductance is replaced by the ratio mean current over voltage. In ii) we show that the discrete description compatible with thermodynamics leads to the continuous one of i) with again the H-K prescription. Here the generalized Johnson-Nyquist relation for noise is recovered only at low voltage, when the continuous description is valid.
	
	
	\vspace{0.5 cm}
	\fontsize{12}{14}\selectfont

	\textbf{Introduction.}
	Models of noise in resistive electrical components must be compatible with thermodynamics, otherwise they lead to unphysical predictions such as the Brillouin paradox \cite{Brillouin50}, an incarnation of Maxwell's demon in electronic systems. So far, two main models have been mostly considered in the classical realm: the Johnson-Nyquist model and the shot noise model \cite{Gillespie2000}. While the former is valid only for linear resistors, the latter approach makes it possible to describe nonlinear components, but at the cost of greater theoretical difficulty.

	In this work, we first consider the possibility of modeling a nonlinear resistor by means of Gaussian white noise. The use of stochastic differential equations (SDE) in the presence of multiplicative noise raises the question of the choice of a prescription, the most common being those of Itô and Stratonovich \cite{vanKampen81}. It has been shown that the use of these prescriptions in the model mentioned above leads to a dead end \cite{Wyatt99}: it seems impossible to conciliate non-linearity and Gaussian white noise. While Itô and Stratonovich prescriptions are commonly used respectively in finance and physics \cite{Kleinertbook}, an infinite choice of prescription is possible, that can be implemented via a continuous parameter $\alpha$ \cite{Cugliandolo22}. Using an arbitrary prescription to the model, we show that there is an issue: one can have a nonlinear resistor that generates Gaussian noise, provided one uses the Hänggi-Klimontovich (H-K) prescription in the SDE. This prescription had already been mentioned in the context of the study of mesoscopic heat engine \cite{Buttiker12}. We find that under this condition there is only one way to generalize the Johnson-Nyquist relation. The relation we find applies to any voltage controlled nonlinear passive resistor (the extension to current controlled case is straightforward).
	
	In a second part, we focus on the discrete, shot noise model for current fluctuations, well described by a master equation \cite{vanKampenbook}. Such an equation involves jump rates between charge states of the circuit. Here again, an arbitrary parameter $\tilde\alpha$ is involved, that describes causality: should the jump rate depend on the initial charge, the final charge, or any intermediate? By taking the continuous voltage limit of the master equation we show how $\alpha$ and $\tilde\alpha$ are related, and in particular that the choice of $\tilde\alpha$ that avoids Brillouin paradox corresponds to the H-K prescription for $\alpha$. Here however, the generalized Johnson-Nyquist formula for noise is recovered only at low voltage.

	\vspace{0.3cm}
	
	\textbf{Continuous, Gaussian Noise Model.} Let us consider a resistive element in parallel with a capacitor $C$. The charge across the capacitor is denoted $ Q_t $ and the voltage $ U_t $. If we model the resistive element using a Gaussian white noise (derived from a Wiener process $ W_t $), the voltage $ U_t $ satisfies a stochastic differential equation (SDE) of the form
	\begin{equation}
		C \dd U_t = - \mathcal{I}(U_t) \dd t - \mathcal{J}(U_t) \dd W_t,  \label{EDS} 
	\end{equation}
	where $ \mathcal{I}(u) $ gives the deterministic $ I(V) $ curve, and $ \mathcal{J}(u) $ describes the multiplicative noise, depending on the bias voltage. For a fixed voltage $u$, $ \mathcal{J}(u) $ represents the spectral density of the standard deviation of the current fluctuations, in A$/\sqrt{\mathrm{Hz}}$. This SDE acquires a precise mathematical definition when a time discretization is given. Such a discretization is not unique, the choice of it being parametrized by a parameter $ \alpha \in [0, 1] $ \cite{Cugliandolo22}. The discretized SDE then takes the following form, written at time steps $ t_n $ and $ t_{n+1} $:    
	\begin{equation}
		C (U_{t_{n+1}} - U_{t_{n}}) = -\mathcal{I}(U_{t_n})(t_{n+1}-t_{n}) - \mathcal{J}\big(U_{t_n} + \alpha(U_{t_{n+1}} - U_{t_n}) \big) ( W_{t_{n+1}} - W_{t_{n}} ).
		\label{SDEd}
	\end{equation}
	The common Itô and Stratonovich prescriptions correspond to $ \alpha = 0 $ and $ 1/2 $  respectively. Once  a choice for $ \alpha $ is made, the discretized SDE \eqref{SDEd} can be associated with a Fokker-Planck equation verified by the probability density $ P(u, t) $: 
	\begin{equation}
		\begin{split}
			\dot P(u, t)  = - \dfrac{\dr }{\dr u} \bigg\{ \bigg[ - \dfrac{\mathcal{I}(u)}{C} + \alpha \dfrac{\mathcal{J}(u)\mathcal{J}'(u)}{C^2}   \bigg] P(u, t)  -  \dfrac{1}{2C^2} \dfrac{\dr }{\dr u} \Big[ \big(  \mathcal{J}(u)\big)^2 P(u, t) \Big]   \bigg\}.
		\end{split}
	\end{equation}
	At thermal equilibrium (in the steady state), the probability current must vanish, $\dot P(u, t)  = 0$:
	\begin{equation}
		\bigg[ - \dfrac{\mathcal{I}(u)}{C} + \alpha \dfrac{\mathcal{J}(u)\mathcal{J}'(u)}{C^2}   \bigg] P_{\mathrm{eq}}(u) -  \dfrac{1}{2C^2} \dfrac{\dr }{\dr u} \Big[ \big(  \mathcal{J}(u)\big)^2 P_{\mathrm{eq}}(u) \Big]  = 0,
	\end{equation}
	and the probability distribution must follow the Maxwell-Boltzmann statistic, i.e. $ P_{\mathrm{eq}}(u) \propto \exp (- Cu^2/2\kb T) $ ($ \kb $ being the Boltzmann constant and $T$ the temperature). The combination of these two conditions leads to the following equation : 
	\begin{equation}
		- \mathcal{I}(u) + \mathcal{J}(u)\mathcal{J}'(u)(\alpha - 1)/C + u \big(\mathcal{J}(u)\big)^2 /(2\kb T) = 0.
		\label{FPeq}
	\end{equation}
	$ \mathcal{I}(u) $ and $ \mathcal{J}(u) $ represent characteristics of the nonlinear resistor only, so they must be independent of $C$. Eq. (\ref{FPeq}) had already been obtained in \cite{Wyatt99} for the particular case of Itô ($\alpha=0$) and Stratonovich ($\alpha=1/2$) prescriptions. In such cases, \eqref{FPeq} implies $ \mathcal{I}(u) = G u $ with a constant conductance $ G > 0 $, and $ \mathcal{J}(u) = \sqrt{2 \kb T G} $: this corresponds to the linear resistor, which noise obeys the usual Johnson-Nyquist relation. Obviously, this property holds for any $ \alpha \neq 1 $. Having extended the approach of \cite{Wyatt99} to any prescription allows us to see that there is another possibility,  $ \alpha = 1 $, directly leading to the generalized Johnson-Nyquist relation
	\begin{equation}
		\mathcal{J}(u) = \sqrt{2 \kb T \mathcal{I}(u)/u},
	\end{equation}
	and to the condition $ \mathcal{I}(u)/u > 0 $ (meaning that the resistor dissipates energy). Taking $ \alpha = 1 $ corresponds to the H-K prescription,  already encountered for modeling fluctuations in a mesoscopic heat-engine \cite{Buttiker12}. Our result shows that this prescription is mandatory to describe a nonlinear resistor with Gaussian noise.

	\vspace{0.3cm}
	\textbf{Brillouin paradox, mechanical analog.} Ignoring the previous discussion and assuming $ \mathcal{J}(u) = \mathrm{constant} $ while keeping a nonlinear $ \mathcal{I}(u) $ implies that $ \langle U_t\rangle_{\mathrm{eq}} \neq 0 $, which corresponds to the well-known Brillouin paradox \cite{Brillouin50}: it implies that one could extract work from a single thermal bath coupled with a nonlinear resistance, in contradiction with the second principle of thermodynamics.
	In a similar way, assuming the validity of the generalized Johnson-Nyquist relation but taking $ \alpha \neq 1 $, also implies in general $ \langle U_t\rangle_{\mathrm{eq}} \neq 0 $. 
	
	A mechanical analog of the nonlinear resistor considered above consists of a particle of mass $m$ evolving at velocity $V_t$, and submitted to a velocity-dependent damping force $ -\gamma(V_t) V_t $ (with $ \gamma(v) \geqslant 0 $), and a fluctuating force $ \Gamma(V_t) \dot W_t $ due to the collisions of the particle with the molecules of the fluid. Applying the fundamental principle of dynamics yields:
	\begin{equation}
		m \dd V_t = - \gamma(V_t) V_t \dd t - \Gamma(V_t) \dd W_t.  \label{EDS2} 
	\end{equation}
	The generalized Johnson-Nyquist relation reads
	\[
	\Gamma(v) = \sqrt{2\kb T \gamma(v) }.
	\]
	In this case, the physical meaning of the Brillouin paradox is transparent: if the generalized Johnson-Nyquist relation is not fulfilled (or, equivalently, if $ \alpha \neq 1 $), then $ \langle V_t\rangle_{\mathrm{eq}} \neq 0 $: the random collisions of the fluid on the particle would give a finite drift velocity to the particle. It would then be possible to extract work from the particle, despite the fact that there is only one thermal bath.

	\vspace{0.3cm}
	
	\textbf{Discrete, Shot Noise Model.} When the noise generated by the resistor is modeled by a Markov jump process, the voltage takes discrete values $ u_n = nv $, with $ n \in \mathbb{Z}$, $v=\qe /C $ and $\qe$ is the electron charge. The probability $P(u_n, t)$ of the voltage being $u_n$ obeys  the following master equation \cite{Wyatt99,Esposito21}:
	\begin{equation}
		\dot{P}(u_n, t)= \lambda_{n\leftarrow n+1 } P(u_{n+1}, t) + \lambda_{n\leftarrow n-1 } P(u_{n-1}, t) - \lambda_{n+1\leftarrow n } P(u_n, t) - \lambda_{n-1\leftarrow n } P(u_n, t).
	\end{equation}
	Here $ \lambda_{p\leftarrow n} $ is the rate for the jump from the charge state $n$ to the charge state $p=n\pm1$, i.e. from voltage $u_n$ to voltage $u_n\pm v$. These rates depend on the voltage, which can be chosen between $u_n$ and $u_p$. We introduce the parameter $ \tilde{\alpha} \in [0, 1] $ which describes this choice, by taking the relevant voltage to be $u_n+\tilde\alpha(u_p-u_n)$. $\tilde\alpha$ describes the causality: $\tilde\alpha=0$ corresponds to the jump depending on the initial state only, while for $\tilde\alpha=1$ only the final state matters. Let us introduce the functions $ \lambda_\pm(u) $ such that: 
	\begin{equation}
		\begin{split}
			\lambda_{n \leftarrow n+1} & = \lambda_-\big( u_{n+1} + \tilde{\alpha} (u_{n} - u_{n+1})  \big) =  \lambda_-( u_{n+1} - \tilde{\alpha} v  ) = \lambda_-\big( u_{n} - (\tilde{\alpha}-1) v  \big)  , \\
			\lambda_{n \leftarrow n-1} & = \lambda_+\big( u_{n-1} + \tilde{\alpha} (u_{n} - u_{n-1})  \big) = \lambda_+( u_{n-1} + \tilde{\alpha} v  )  = \lambda_+\big( u_{n} + (\tilde{\alpha}-1) v  \big), \\
			\lambda_{n+1 \leftarrow n} & = \lambda_+\big( u_n + \tilde{\alpha} (u_{n+1} - u_n)  \big) = \lambda_+( u_n + \tilde{\alpha} v  ), \\
			\lambda_{n-1 \leftarrow n} & = \lambda_-\big( u_{n} + \tilde{\alpha} (u_{n-1} - u_{n})  \big) =  \lambda_-( u_{n} - \tilde{\alpha} v  ),
		\end{split}
	\end{equation}
	Writing $ u = u_n $ yields the desired equation:    
	\begin{equation}
		\begin{split}
			\dot P(u, t) & = \lambda_-\big( u - (\tilde{\alpha}-1) v  \big) P(u + v, t) + \lambda_+\big( u + (\tilde{\alpha}-1) v  \big) P(u - v, t)  \\
			& \qquad - \Big[ \lambda_+( u + \tilde{\alpha} v , t ) +\lambda_- ( u - \tilde{\alpha} v  )   \Big] P(u, t).
		\end{split}
		\label{Meq}
	\end{equation}
	The functions $\lambda_\pm(u)$ satisfy the Local Detailed Balance condition \cite{Esposito21,Maes21}, $ \lambda_+(u)/\lambda_-(u) = \ee^{\qe u /\kb T} $. When the average current is not a linear function of the voltage $u$, taking $ \tilde{\alpha} = 1/2 $ is necessary to avoid the Brillouin paradox \cite{Wyatt99, Esposito21}.  Expanding equation \eqref{Meq} up to order two in $v$ gives a Fokker-Planck equation with the prescription $ \alpha = 2 \tilde{\alpha} $:
	\begin{equation}
		\begin{split}
			\dot P(u, t) & = \Big[  v \big( \lambda'_-(u) - \lambda'_+(u)  \big) + \dfrac{1 - 2\tilde{\alpha}}{2} v^2 \big( \lambda_+''(u) + \lambda''_-(u)  \big)   \Big] P(u, t)  \\
			& \quad + \Big[ v \big( \lambda_-(u) - \lambda_+(u)  \big)  + v^2 (1 - \tilde{\alpha}) \big( \lambda'_+(u) + \lambda'_-(u) \big)   \Big]  P'(u, t)  \\
			& \quad + \dfrac{v^2}{2} \big( \lambda_+(u) + \lambda_-(u) \big) P''(u, t).
		\end{split}
		\label{FP}
	\end{equation}
	This equation is acceptable only for $ \tilde{\alpha} \in [0, 1/2] $. For larger values of $\tilde\alpha$ no Fokker-Planck equation can be associated to the master equation \eqref{Meq}. The correspondence $ \alpha = 2 \tilde{\alpha} $  is somehow remarkable: while $ \alpha $, as a prescription parameter for an SDE, is directly related to the discretization of time, $ \tilde{\alpha} $ is, on the contrary, present in the master equation describing an evolution in continuous time but discrete in charge. The value $ \tilde{\alpha} = 1/2 $ corresponds to $ \alpha = 1 $, i.e. to the H-K prescription. Once again, the H-K condition appears as the only physically acceptable prescription in the limit $ |\qe u | \ll \kb T $. Indeed, from Eq.\eqref{FP}, the noise amplitude is given by $ \mathcal{J}(u) = \sqrt{\qe\mathcal{I}(u)\coth(\qe u/2k_BT)} $ with $ \mathcal{I}(u)=\qe[\lambda_+(u) - \lambda_-(u)]  $. The generalized Johnson-Nyquist relation is thus verified only, but universally, at low voltage, that is when $|\qe u|\ll \kb T$. This last condition ensures that the Local Detailed Balance condition is verified up to order two in $ \qe u/\kb T $. Indeed, taking $ \mathcal{J}(u) = \sqrt{2\kb T\mathcal{I}(u)/u} $ yields:
	\begin{equation}
		\dfrac{\lambda_+(u)}{\lambda_-(u)} = \dfrac{1 + \qe u/2\kb T  }{1 - \qe u/2\kb T} = 1 + \qe u/\kb T + \dfrac{1}{2} (\qe u/\kb T)^2 + ... .
	\end{equation}

	\vspace{0.3cm}
	
	\textbf{Conclusion and perspectives.} 
	We have shown that a nonlinear resistor having Gaussian noise is possible according to thermodynamics, provided its noise spectral density is related to its $I(V)$ characteristics by a generalized Johnson-Nyquist relation. In terms of stochastic differential equations, this is accompanied by the necessity to use the H-K prescription. Furthermore, taking the continuous limit of the shot noise model, we have also shown that the recipe used to go from the discrete charge to the continuous charge limit while avoiding the Brillouin paradox corresponds, in fact, to the H-K prescription in terms of SDE. In this limit, the generalized Johnson-Nyquist formula is recovered only at low voltage. 
	
	The results of the present work and their limitations raise several questions. We only considered classical fluctuations, but at sufficiently low temperatures it is necessary to take into account quantum effects. Furthermore, the discussion involved a unique nonlinear resistor in contact with a unique degree of freedom. It is known that when an electromagnetic environment is present (modeled by an impedance), retroactive effects occur \cite{Ingold1992,Karl21}. The extension of the present results for circuits comprising several (nonlinear) resistive elements in which these retroactive effects occur is left for future work.

	\vspace{0.3cm}
	
	\textbf{Acknowledgments.} We would like to thank Léopold Van Brandt and Jean-Charles Delvenne for fruitful discussions. This work was supported by the Canada Research Chair program, the NSERC, the Canada First Research Excellence Fund, the FRQNT, and the Canada Foundation for Innovation.
	
	\vspace{0.3cm}

	\bibliography{Biblio}
	\bibliographystyle{unsrt}


\end{document}